\documentclass[aps,prb,showpacs]{revtex4}
\usepackage{latexsym}
\usepackage{amssymb}
\usepackage{graphicx}
\begin{document}
\title{Spin Hall effect in infinitely large and finite-size diffusive Rashba two-dimensional electron systems:
A helicity-basis nonequilibrium Green's function approach}
\author{S. Y. Liu}
\email{liusy@mail.sjtu.edu.cn}
\author{X. L. Lei}
\affiliation{Department of Physics, Shanghai Jiaotong University, 1954
Huashan Road, Shanghai 200030, China}
\date{\today}
\begin{abstract}
A nonequilibrium Green's function approach is employed to
investigate the spin-Hall effect in diffusive two-dimensional
electron systems with Rashba spin-orbit interaction. Considering a
long-range electron-impurity scattering potential in the
self-consistent Born approximation, we find that the spin-Hall
effect arises from two distinct interband polarizations in
helicity basis: a disorder-unrelated polarization directly induced
by the electric field and a polarization mediated by
electron-impurity scattering. The disorder-unrelated polarization
is associated with all electron states below the Fermi surface and
produces the original intrinsic spin-Hall current, while the
disorder-mediated polarization emerges with contribution from the
electron states near the Fermi surface and gives rise to an
additional contribution to the spin-Hall current. Within the
diffusive regime, the total spin-Hall conductivity vanishes in
{\it infinitely large} samples, independently of temperature, of
the spin-orbit coupling constant, of the impurity density, and of
the specific form of the electron-impurity scattering potential.
However, in a {\it finite-size} Rashba two-dimensional
semiconductor, the spin-Hall conductivity no longer always
vanishes. Depending on the sample size in the micrometer range, it
can be positive, zero or negative with a maximum absolute value
reaching as large as $e/8\pi$ order of magnitude at low
temperatures. As the sample size increases, the total spin-Hall
conductivity oscillates with a decreasing amplitude. We also
discuss the temperature dependence of the spin-Hall conductivity
for different sample sizes.

\end{abstract}

\pacs{72.10.-d, 72.25.Dc, 73.50.Bk}
\maketitle

\section{Introduction}
The proposed spin-Hall effect, namely, the appearance of a spin
current along the direction perpendicular to the driving electric
field, has attracted much recent theoretical and experimental
attention. In early studies, it was shown that an {\it extrinsic}
spin-Hall effect may arise from a spin-orbit (SO) interaction of
electrons induced by electron-impurity scattering
potential.\cite{DP,HS} Recently, a scattering-independent {\it
intrinsic} spin-Hall effect, which entirely originates from a
spin-orbit coupling in the free-carrier system itself, has been
predicted respectively in $p$-type bulk semiconductors\cite{t1}
and $n$-type two-dimensional (2D) systems with Rashba\cite{t2} and
Dresselhaus SO interaction.\cite{t3} The experimental observations
of the spin-Hall effect have been also reported in a $n$-type bulk
semiconductor\cite{Kato} and in a two-dimensional heavy-hole
system.\cite{Wunderlich}

In two-dimensional electron systems with Rashba SO interaction,
ignoring the effect of disorders, Sinova {\it et al.} showed that
the spin-Hall conductivity, $\sigma_{sH}$, has a universal
intrinsic value $e/8\pi$ at zero temperature.\cite{t2}
Subsequently, a great deal of research work was focused on the
influence of disorders on this intrinsic spin-Hall effect. When
Rashba two-dimensional electron systems are sufficiently dirty and
the Anderson localization is dominant, Sheng {\it et al.} found
that the spin-Hall conductivity can be much greater or smaller
than the universal value $e/8\pi$.\cite{t4} In the diffusive
regime, it was demonstrated that the collisional broadening in the
density of states of electrons leads to a reduction of the
spin-Hall current and the $e/8\pi$ value of $\sigma_{sH}$ can be
reached only for relatively weak electron-impurity
scattering.\cite{Nomura,t6} However, further investigation
revealed that the electron-impurity scattering can also produce an
additional contribution to spin-Hall current, which is independent
of the impurity density and has a sign opposite to the original
one. As a result, the total spin-Hall current is completely
suppressed for a short-range electron-impurity scattering
potential.\cite{t8} This conclusion has been confirmed by
different methods, such as Kubo formula,\cite{t8,t7,t9} Keldysh
formalism,\cite{t88} and spin-density method\cite{t10} {\it etc}.
Also, it made clear that this cancellation of spin-Hall current is
not due to any symmetry.\cite{Murakami}

However, in most previous studies, the vanishing of the spin-Hall
current was found only for a short-range electron-impurity
scattering.\cite{Nomura,t6,t7,t9,t8,t88,t10} It is well known
that, in realistic 2D semiconductor systems, the dominant
electron-impurity collisions are long-ranged. It is interesting to
see whether the spin-Hall current survives in the case of
long-range electron-impurity scattering. In Ref.\,\onlinecite{t8},
the authors argued that the total spin-Hall current should be {\it
nonvanishing} in the case of long-range electron-impurity
collisions. However, Raimondi and Schwab again got a vanishing
spin-Hall conductivity considering a weakly momentum-dependent
potential: this potential depends on the cosine of the angle
between the initial and the final momenta, but is independent of
their magnitudes.\cite{Raimondi}

In this paper, we carefully investigate the spin-Hall current in
2D electron systems with Rashba SO coupling by means of a
nonequilibrium Green's function approach. We consider a quite
general form of the electron-impurity scattering potential: it
depends not only on the directions but also on the magnitudes of
the electron momenta. Such a potential can be used to describe the
realistic Coulomb interaction between electrons and impurities in
2D semiconductors. Besides, in contrast to all of the previous
discussions concerning electron behaviors in the spin
basis,\cite{Nomura,t6,t7,t9,t8,t88,t10,Raimondi} our
formalism is presented in the helicity basis. Such a treatment
allows us to interpret the origin of the spin-Hall effect in terms
of interband polarization processes. We clarify that the spin-Hall
current arises from two mechanisms: disorder-unrelated and
disorder-mediated mechanisms, which correspond to two distinct
helicity-basis interband polarizations. The disorder-unrelated
mechanism is associated with a polarization directly induced by dc
electric field, which results in the original intrinsic spin-Hall
current with contribution from all electron states in the Fermi
sea. The disorder-mediated mechanism relates to a polarization
mediated by electron-impurity scattering and is associated mainly
with the electron states in the vicinity of the Fermi surface.
We find that in infinitely large 2D Rashba
semiconductors, the total spin-Hall current vanishes,
independently of the specific form of the electron-impurity scattering
potential, of the impurity density, of the SO coupling constant,
and of temperature.

However, we also make clear that the "always vanishing" of the
spin-Hall current occurs only for {\it infinitely large} samples.
Care must be taken in regard with this conclusion when the sample
size reduces. The discretization of the energy levels in
finite-size system may lead to a nonvanishing total spin-Hall
current in {\it finite} Rashba 2D semiconductors even in the
quasiclassical regime. Numerical calculation for square shape
Rashba 2D electron systems of size in micrometer regime, indicates
that depending on the system size, the total spin-Hall
conductivity can be positive, zero or negative, with a maximum
absolute value reaching up to the order of magnitude of $e/8\pi$
at low temperatures. As a function of the sample size,
$\sigma_{sH}$ oscillates around zero with a decreasing amplitude
when increasing sample size. Such a size effect can be observable
only at low temperatures. When temperature increases that $T$
becomes comparable with the finite-size induced energy separation
of the electron states at the Fermi surface, $\sigma_{sH}$
oscillation disappears and the spin-Hall conductivity returns to a
small nonvanishing value before it slowly approaches zero with
further increasing sample size.

This paper is organized as follows. In Sec. II, we present a
general formalism for the nonequilibrium Green's functions. Based
on this, the mechanisms of the spin-Hall effect are clarified. In
Sec. III, we give analytical and numerical calculations of the
spin-Hall conductivity in both the infinitely large and
finite-size Rashba 2D semiconductors. Finally, we review our
results in Sec. IV.

\section{General formalism}
\subsection{Kinetic equation for lesser Green's function}
We consider a quasi-2D electron semiconductor in $x-y$ plane,
subjected to a Rashba SO interaction. The single-particle
noninteracting Hamiltonian of the system can be written as
\begin{equation}
\check h = \frac{{\bf p}^2}{2 m}+ \alpha {\bf p} \cdot ({\bf n}
\times{\hat {\bf \sigma}}  ),\label{EH}
\end{equation}
where $\alpha$ is the Rashba SO coupling constant, ${\hat{\bf
\sigma}}\equiv ({\hat \sigma}_x, {\hat \sigma}_y, {\hat
\sigma}_z)$ are the Pauli matrices, $m$ is the electron effective
mass, ${\bf p}\equiv (p_x,p_y)\equiv (p\cos\phi_{\bf
p},p\sin\phi_{\bf p})$ is the 2D electron momentum, and ${\bf n}$
is the unit vector perpendicular to the 2D electron plane.
This Hamiltonian can be diagonalized,\cite{t9}
resulting in two eigenvalues $\varepsilon_{\mu}(p)={p}^2/2m+(-1)^\mu\alpha p$
 and eigen wave functions $\varphi_{\mu} ({\bf p})=u_{\mu}({\bf p}){\rm e}^{i{\bf p}\cdot {\bf r}}$
with
\begin{equation}
u_\mu ({\bf p}) = \frac{1}{\sqrt{2}} \left(
\begin{array}{c}
    1\\
    (-1)^{\mu +1}i{\rm e}^{i \phi_{\bf p}}
  \end{array}
 \right),
\end{equation}
and $\mu=1,2$. It is useful to introduce a unitary transformation
$U_{\bf p}=[u_1({\bf p}),u_2({\bf p})]$, by which the basis of the
system is transformed from a spin basis to a helicity basis.

In 2D systems, the electrons experience scattering by impurities.
The previous studies were concerned only with a short-range
interaction between electrons and impurities, corresponding to a
potential independent of electron momentum. However, in realistic
2D semiconductors, such as heterojunctions and quantum wells {\it
etc}, the potential of the Coulomb interaction between the
electrons and impurities essentially is long-ranged. In this
paper, we assume that the electron-impurity scattering can be
described by an isotropic potential $V(|{\bf p}-{\bf k}|)$, which
corresponds to scattering an electron from momentum state ${\bf
p}$ to state ${\bf k}$. This potential depends not only on the
angle $\phi_{\bf p}-\phi_{\bf k}$ but also on the magnitudes of
the momenta ${\bf p}$ and ${\bf k}$. The latter dependence of the
potential has been ignored in Ref.\,\onlinecite{Raimondi}. In
helicity basis, the scattering potential takes a transformed form,
$\hat{T}({\bf p},{\bf k})=\hat U^+({\bf p})V(|{\bf p}-{\bf
k}|)\hat U({\bf k})$.

We are interested in the spin-Hall current driven by a dc electric
field ${\bf E}$ along the $x$ axis. In Coulomb gauge, this
electric field can be described by a scalar potential $V=-e{\bf
E}\cdot {\bf r}$, with ${\bf r}$ as the electron coordinate. In
Rashba 2D electron systems driven by the electric field ${\bf E}$,
the only nonvanishing component of the spin-Hall current is just
the spin-Hall current polarized along the $z$-direction and flow
along the $y$ axis, $J_y^z$. Its single-particle operator, defined
in the spin basis as $\check j_y^z=( \check j_y{\hat
\sigma}_z+{\hat \sigma}_z\check j_y)/4e$ with the electric current
operator $\check j_y$,\cite{RA} reduces to an off-diagonal matrix
in the helicity basis. Taking a statistical ensemble average, the
net spin-Hall current can be determined in the helicity basis via
\begin{equation}
{J}^{z}_y=-i\sum_{{\bf p}}\frac {p_y}{2m}\int \frac{d\omega}{2\pi}
\left [{\hat {\rm G}}^<_{12}
({\bf p},\omega)+{\hat {\rm G}}^<_{21}({\bf p},\omega)\right ]
=\sum_{{\bf p}}\frac {p_y}{m}\int \frac{d\omega}{2\pi}
{\rm Im}{\hat {\rm G}}^<_{12}
({\bf p},\omega),\label{Jz}
\end{equation}
with $\hat {\rm G}^<({\bf p},\omega)$ as the helicity-basis
nonequilibrium lesser Green's function. We see that the spin-Hall
effect in Rashba 2D electron systems arises only from interband
polarization processes. The contribution to $J_y^z$ from the
diagonal elements of $\hat {\rm G}^<({\bf p},\omega)$ vanishes
because $U^+({\bf p})\hat \sigma_z U({\bf p})$ is an off-diagonal
matrix.

In order to investigate the spin-Hall effect, it is necessary to
study the nonequilibrium lesser Green's function $\hat {\rm
G}^{<}_{\bf p}$. For brevity, hereafter, we employ a subscript
${\bf p}$ to denote the arguments of the Green's functions and
self-energies, $({\bf p},\omega)$. In Rashba 2D electron systems
with short-range disorders, the kinetic equation for Keldysh
function, which simply relates to the lesser Green's function, has
already been constructed in the spin basis.\cite{t88} However,
from Eq.\,(\ref{Jz}), we see that it is most convenient to study
$\hat {\rm G}^{<}_{\bf p}$ in the helicity basis. In this basis,
the {\it noninteracting} retarded and advanced Green's functions
$\hat {\rm g}_{\bf p}^{r,a}$ are diagonal, $\hat{ {\rm
g}}^{r,a}_{\bf p}={\rm diag}\left ((\omega-\varepsilon_1(p)\pm
i\delta)^{-1}, (\omega-\varepsilon_2(p)\pm i\delta)^{-1}\right)$,
as well as the {\it interacting} unperturbed ones $\hat {\rm
G}_{0{\bf p}}^{r,a}$: $\hat {\rm G}_{0{\bf p}}^{r,a}=[1+\hat{{\rm
g}}^{r,a}_{\bf p} \hat {\Sigma}_{0{\bf p}}^{r,a}]^{-1}\hat{{\rm
g}}^{r,a}_{\bf p}$ with diagonal unperturbed self-energies $\hat
{\Sigma}_{0{\bf p}}^{r,a}$.

Under steady and homogeneous conditions, the kinetic equation for
the helicity-basis lesser Green's function $\hat {\rm G}^<_{\bf p}$ reads
\begin{equation}
ie {\bf {\rm E}} \cdot \left ({\bf\nabla}_{\bf p} \hat{\rm G}^<_{\bf p} +
\frac {i{\bf \nabla}_{\bf p}\phi_{\bf p}}{2}[\hat{\rm G}^<_{\bf p},{\hat
\sigma}_x] \right ) -\alpha  p [ \hat{\rm G}^<_{\bf p}, {\hat \sigma}_z]
=( \hat{\Sigma}^r_{\bf p}
   \hat{\rm G}^<_{\bf p} - \hat{\rm G}^<_{\bf p} \hat{\Sigma}^a_{\bf p}
   - \hat{\rm G}^r_{\bf p} \hat{\Sigma}^<_{\bf p} +
   \hat{\Sigma}^<_{\bf p} \hat{\rm G}^a_{\bf p} ),\label{EK}
\end{equation}
where, $\hat{\rm G}^{r,a,<}_{\bf p}$ and
$\hat{\Sigma}^{r,a,<}_{\bf p}$, respectively, are the
nonequilibrium Green's functions and self-energies.
Eq.\,(\ref{EK}) is derived from the kinetic equation in the spin
basis by application of the local unitary transformation $U({\bf
p})$. At the same time, in this, only the lowest order of gradient
expansion is taken into account.\cite{JH}

In present paper, we consider the electron-impurity scattering in
the self-consistent Born approximation. It is widely accepted that
this is sufficiently accurate to analyze the transport properties
in diffusive regime. Accordingly, the self-energies take the forms
\begin{equation}
\hat{\Sigma}^{r,a,<}_{\bf p}=n_i\sum_{{\bf k}}\hat T({\bf
p},{\bf k})\hat{\rm G}^{r,a,<}_{\bf k}\hat T^+({\bf p},{\bf
k}),\label{SE}
\end{equation}
with impurity density $n_i$. Substituting explicit form of the
matrix ${U}({\bf p})$ into Eq.\,(\ref{SE}), we get
\begin{equation}
{\hat \Sigma}^{r,a,<}_{\bf p}=\frac 12 n_i\sum_{{\bf
k}}|V(|{\bf p}-{\bf k}|)|^2 \left \{ a_1 \hat {\rm G}^{r,a,<}_{\bf p}
+a_2{\hat \sigma}_x\hat {\rm G}^{r,a,<}_{\bf p}{\hat \sigma}_x+ia_3[{\hat
\sigma}_x,\hat {\rm G}^{r,a,<}_{\bf p}]\right \}.\label{SER}
\end{equation}
Here $a_i$($i=1,2,3$) are the factors associated with the
directions of momenta, $a_1=1+\cos (\phi_{\bf p}-\phi_{\bf k})$,
$a_2=1-\cos (\phi_{\bf p}-\phi_{\bf k})$, $a_3=\sin (\phi_{\bf
p}-\phi_{\bf k})$.

Further, our considerations are restricted to the linear response
regime. In connection with this, all the functions, such as the
nonequilibrium Green's functions and self-energies, can be
expressed as sums of two terms: $A=A_0+A_1$, with $A$ as the
Green's functions or self-energies. $A_0$ and $A_1$, respectively,
are the unperturbed part and the linear electric field part of
$A$. In this way, the kinetic equation for $\hat {\rm G}^<_{1{\bf
p}}$ can be written as
\begin{equation}
-\alpha p \hat{C}_{1{\bf p}}+ ie{\rm \bf E}\cdot  {\bf \nabla}_{\bf
p}\hat{\rm G}_{0{\bf p}}^<-\frac{1}{2} e {\bf \rm E} \cdot{\bf \nabla}_{\bf
p}\phi_{\bf p} \hat{D}_{0{\bf p}}=
 \hat{\Sigma}^r_{0{\bf p}}
   \hat{\rm G}^<_{1{\bf p}} - \hat{\rm G}^<_{1{\bf p}} \hat{\Sigma}^a_{0{\bf p}} - \hat{\rm G}^r_{0{\bf p}}
   \hat{\Sigma}^<_{1{\bf p}} +
   \hat{\Sigma}^<_{1{\bf p}} \hat{\rm G}^a_{0{\bf p}},\label{EQOne}
\end{equation}
where the matrices $\hat{C}_{1{\bf p}}$ and $\hat{D}_{0{\bf p}}$, respectively, are
\begin{equation}
\hat{C}_{1{\bf p}}=\left (
\begin{array}{cc}
0&-2 (\hat{\rm G}^<_{1{\bf p}})_{12}\\
2 (\hat{\rm G}^<_{1{\bf p}})_{21}&0\\
\end{array}
\right ),
\end{equation}
and
\begin{equation}
\hat{D}_{0{\bf p}}=\left (
\begin{array}{cc}
0&(\hat{\rm G}^<_{0{\bf p}})_{11}-(\hat{\rm G}^<_{0{\bf p}})_{22}\\
(\hat{\rm G}^<_{0{\bf p}})_{22}-(\hat{\rm G}^<_{0{\bf p}})_{11}&0
\end{array}
\right ),\label{EG}
\end{equation}
and $\hat{\rm G}^<_{0{\bf p}}$ is the unperturbed lesser Green's
function, $\hat{\rm G}^<_{0{\bf p}}=-2in_{\rm F}(\omega){\rm
Im}\hat{\rm G}^<_{0{\bf p}}$, with $n_{\rm F}(\omega)$ as the
Fermi function. Here, to derive Eq.\,(\ref{EQOne}), we have
employed the vanishing of the contributions to spin-Hall current
from $\hat{\rm G}^{r,a}_{1{\bf p}}$ and $\hat{\Sigma}^{r,a}_{1{\bf
p}}$ involved in the terms on the right-hand side of
Eq.\,(\ref{EK}), which can be easily demonstrated by considering
Eqs.\,(\ref{Jz}) and (\ref{SE}), as well as Eq.\,(\ref{TT})
presented below.

It is obvious that the driving force in Eq.\,(\ref{EQOne})
comprises two components: $ie{\rm\bf E}\cdot {\bf \nabla}_{\bf
p}\hat{\rm G}_{0{\bf p}}^<$ and $ - e {\rm\bf  E} \cdot{\bf
\nabla}_{\bf p}\phi_{\bf p} \hat{D}_{0{\bf p}}/2$. Due to the
linearity of Eq.\,(\ref{EQOne}) that its solution can be assumed
to be a sum of two terms $(\hat {\rm G}_{1{\bf p}}^{<})^{I}$ and
$(\hat{\rm G}_{1{\bf p}}^{<})^{II}$, which, respectively, obey the
following equations,
\begin{equation}
-\alpha p \hat{C}_{1{\bf p}}^{I}+ie{\rm \bf E}\cdot  {\bf \nabla}_{\bf
p}\hat{\rm G}_{0{\bf p}}^<= \hat{\Sigma}^r_{0{\bf p}}
   (\hat{\rm G}^<_{1{\bf p}})^{I} - (\hat{\rm G}^<_{1{\bf p}})^{I} \hat{\Sigma}^a_{0{\bf p}}
   - \hat{\rm G}^r_{0{\bf p}} (\hat{\Sigma}^<_{1{\bf p}})^{I} +
   (\hat{\Sigma}^<_{1{\bf p}})^{I} \hat{\rm G}^a_{0{\bf p}},\label{EQTwo}
\end{equation}
\begin{equation}
-\alpha p \hat{C}_{1{\bf p}}^{II}-\frac{1}{2} e {\bf \rm E} \cdot{\bf
\nabla}_{\bf p}\phi_{\bf p} \hat{D}_{0{\bf p}}= \hat{\Sigma}^r_{0{\bf p}}
   (\hat{\rm G}^<_{1{\bf p}})^{II} - (\hat{\rm G}^<_{1{\bf p}})^{II} \hat{\Sigma}^a_{0{\bf p}}
   - \hat{\rm G}^r_{0{\bf p}} (\hat{\Sigma}^<_{1{\bf p}})^{II} +
   (\hat{\Sigma}^<_{1{\bf p}})^{II} \hat{\rm G}^a_{0{\bf p}}.\label{EQThree}
\end{equation}
Here, $(\hat{\Sigma}^<_{1{\bf p}})^{I}$ and
$(\hat{\Sigma}^<_{1{\bf p}})^{II}$ are the corresponding
self-energies, corresponding to the Green's functions $(\hat{\rm
G}^<_{1{\bf p}})^{I}$ and $(\hat{\rm G}^<_{1{\bf p}})^{II}$,
respectively.

\subsection{disorder-unrelated mechanism of the spin-Hall effect}

The solution of Eq.\,(\ref{EQThree}) is off-diagonal and can be
derived analytically,
\begin{equation}
(\hat{\rm G}_{1{\bf p}}^<)^{II}_{12}=(\hat{\rm G}^{<}_{1{\bf p}})^{II}_{21}=\frac
{ieE}{2 \alpha p^2} \sin \phi_{\bf p} n_{\rm F}(\omega){\rm Im}
[({{\hat {\rm G}}_{0{\bf p}}}^r)_{11}-({{\hat {\rm
G}}_{0{\bf p}}}^r)_{22}].\label{R1}
\end{equation}
Substituting Eq.\,(\ref{R1}) into Eq.\,({\ref{Jz}}), we obtain the
contribution from $(\hat{\rm G}_{1{\bf p}}^<)^{II}$ to the
spin-Hall conductivity:
\begin{equation}
\sigma_{sH}^{II}=\frac{-e}{4 m\alpha}\sum_{{\bf p}}
\frac{p_y^2}{p^3} [f_{1}(p)-f_{2}(p)],\label{SSH1}
\end{equation}
with ($\mu=1,2$)
\begin{equation}
f_{\mu}(p)=-2\int \frac{d\omega}{2\pi}n_{\rm F} (\omega)
{\rm Im} (\hat {\rm G}^r_{0{\bf p}})_{\mu\mu}
\end{equation}
as the unperturbed distribution function.

$\sigma_{sH}^{II}$ arises from the off-diagonal driving force, $ -
e {\rm\bf E} \cdot{\bf \nabla}_{\bf p}\phi_{\bf p} D_{0{\bf
p}}/2$, which is associated with the electric dipole matrix
\begin{equation}
-\int d{\bf r}\varphi^*_{\mu}({\bf p})e{\bf E}\cdot {\bf r}
\varphi_{\nu}({\bf p}')= -ie{\bf E}\cdot \frac{\partial}{\partial
{\bf p}}\int d{\bf r}\varphi^*_{\mu}({\bf p}) \varphi_{\nu}({\bf
p}') +ie{\bf E}\cdot\frac{\partial}{\partial {\bf
p}}u^*_{\mu}({\bf p})u^*_{\nu}({\bf p}') \int d{\bf r}{\rm
e}^{i({\bf p}'-{\bf p})\cdot {\bf r}}.\label{DM}
\end{equation}
To the first order of the electric field, the off-diagonal
elements of the electric dipole moment ({\it i.e.} the second term
on the right hand side of Eq.\,(\ref{DM})) describes a
polarization process between different bands, directly induced by
the dc electric field. Hence, in essence, $\sigma_{sH}^{II}$
originates from this polarization process and becomes
disorder-unrelated, although it may depend on the scattering
through the collisional broadening in $\hat {\rm G}^r_{0{\bf p}}$.

We note that the polarization process directly induced by the dc
electric field is not restricted to the electron states near
the Fermi surface: it comes from all electron states in the
Fermi sea. As a result, $\sigma_{sH}^{II}$ depends on the
distribution function $f_\mu(p)$ itself, rather than its derivative.

\subsection{Disorder-mediated mechanism of the spin-Hall effect}

The off-diagonal element of solution of Eq.\,(\ref{EQTwo}), $(\hat
{\rm G}_{1{\bf p}}^<)^I_{12}$, can be formally expressed  as
\begin{equation}
(\hat {\rm G}_{1{\bf p}}^<)^I_{12}=\frac 1{2\alpha p} \hat
I_{12},\label{EQG1}
\end{equation}
where $\hat I$ is the term on the right hand side of
Eq.\,(\ref{EQTwo}):
\begin{equation}
\hat I_{\mu\mu}=2i[{\rm Im} ({\hat \Sigma}_{0{\bf p}}^r)_{\mu\mu}
(\hat{\rm G}_{1{\bf p}}^<)^I_{\mu\mu}- {\rm Im} ({\hat {\rm
G}}_{0{\bf p}}^r)_{\mu\mu} (\hat{\Sigma}_{1{\bf
p}}^<)^I_{\mu\mu}],
\end{equation}
\begin{equation}
\hat I_{\mu\bar\mu}=[({\hat \Sigma}_{0{\bf p}}^r)_{\mu\mu}-({\hat
\Sigma}_{0{\bf p}}^a)_{\bar\mu\bar\mu}] (\hat{\rm G}_{1{\bf
p}}^<)^I_{\mu\bar\mu}
 - [({\hat {\rm
G}}_{0{\bf p}}^r)_{\mu\mu}-({\hat {\rm G}}_{0{\bf
p}}^a)_{\bar\mu\bar\mu}] (\hat{\Sigma}_{1{\bf
p}}^<)^I_{\mu\bar\mu},
\end{equation}
with $\bar \mu=3-\mu$. Substituting Eq.\,(\ref{EQG1}) into
Eq.\,(\ref{Jz}), the contribution to  spin-Hall current from
$(\hat {\rm G}_{1{\bf p}}^<)^I$, $\left . J_y^z\right |^I$, is
given by
\begin{equation}
\left . J_y^z\right |^I=\sum_{\bf p}\int \frac{d\omega}{2\pi}\frac{p_y}{2m\alpha p}\left \{ {\rm Re}(\hat
{\rm G}_{1{\bf p}}^<)^I_{12}[{\rm Im} (\Sigma_{0{\bf
p}}^r)_{11}+{\rm Im} (\Sigma_{0{\bf p}}^r)_{22}]
 +{\rm Im}(\hat {\rm G}_{1{\bf
p}}^<)^I_{12}[{\rm Re} (\Sigma_{0{\bf p}}^r)_{11}+{\rm Re}
(\Sigma_{0{\bf p}}^r)_{22}] -(\hat\Sigma\leftrightarrow\hat{\rm
G}) \right \}.\label{B4}
\end{equation}
By inserting the explicit forms of the self-energies, it can be
further simplified as
\begin{eqnarray}
\left . J_y^z\right |^I&=&\frac{1}{2m\alpha}\sum_{{\bf p},{\bf k}}\int \frac{d\omega}{2\pi}|V({\bf p}-{\bf k})|^2
\cos\phi_{\bf k}\sin(\phi_{\bf p}-\phi_{\bf k}){\rm Re}(\hat{\rm
G}_{1{\bf p}}^<)^I_{12} [{\rm Im} (\hat {\rm G}_{0{\bf
k}}^r)_{11}+{\rm Im} (\hat {\rm G}_{0{\bf k}}^r)_{22}]
\nonumber\\
&&-\frac 1{4m\alpha}\sum_{{\bf p},{\bf k}}|V({\bf p}-{\bf
k})|^2[\cos\phi_{\bf k}-\cos\phi_{\bf p}\cos(\phi_{\bf
p}-\phi_{\bf k})] [{\rm Im}(\hat {\rm G}_{0{\bf p}}^r)_{11}+{\rm
Im}(\hat {\rm G}_{0{\bf p}}^r)_{22}] [{\rm Im}(\hat {\rm G}_{1{\bf
k}}^<)_{11}-{\rm Im}(\hat {\rm G}_{1{\bf k}}^<)_{22}].\label{SIT}
\end{eqnarray}
Here, we have considered the vanishing of the real parts of the
diagonal elements of $(\hat {\rm G}_{1{\bf p}}^<)^I$ according to
its definition. Also we have used the relation
\begin{equation}
\sum_{\bf k}|V({\bf
p}-{\bf k})|^2\sin(\phi_{\bf p}-\phi_{\bf k})[(\hat {\rm
G}^r_{0{\bf k}})_{22} -(\hat {\rm G}^r_{0{\bf
k}})_{11}]=0,\label{TT}
\end{equation}
which is derived from angular independence of $\hat {\rm
G}^r_{0{\bf p}}$. Combining the terms proportional to
$\cos\phi_{\bf p}\cos(\phi_{\bf p}-\phi_{\bf k})$ in the second
line of Eq.\,(\ref{SIT}) with a similar term in the first line, we
obtain
\begin{eqnarray}
\left . J_y^z\right |^I&=&\frac 1{4m\alpha} \sum_{{\bf p}}\cos\phi_{\bf p}\left
\{[{\rm Im}(\hat \Sigma_{1{\bf p}}^<)^I_{11} -{\rm
Im}(\hat\Sigma_{1{\bf p}}^<)^I_{22}] [{\rm Im} (\hat {\rm
G}_{0{\bf p}}^r)_{11}+{\rm Im}
(\hat {\rm G}_{0{\bf p}}^r)_{22}]\right .\nonumber\\
&& \left. -[{\rm Im}(\hat\Sigma_{0{\bf p}}^r)^I_{11}+ {\rm
Im}(\hat\Sigma_{0{\bf p}}^r)_{22}] [{\rm Im} (\hat {\rm G}_{1{\bf
p}}^<)_{11}-{\rm Im}(\hat {\rm G}_{1{\bf p}}^<)^I_{22}]\right
\}.\label{CAI}
\end{eqnarray}
Further, we note that in the self-consistent Born approximation,
there is a vanishing quantity
\begin{equation}
{\cal K}\equiv \frac{1}{4m\alpha}\sum_{{\bf p}}\cos \phi_{\bf p}\left \{[{\rm Im}
(\hat {\rm G}_{0{\bf p}}^r)_{11}-{\rm Im} (\hat {\rm G}_{0{\bf
p}}^r)_{22}] [{\rm Im}(\hat \Sigma_{1{\bf p}}^<)^I_{11}+{\rm
Im}(\hat\Sigma_{1{\bf p}}^<)^I_{22}] -(\Sigma\leftrightarrow {\rm
G})\right \}.\label{K}
\end{equation}
The fact of ${\cal K}=0$ can be shown by inserting the
explicit forms of the self-energies, Eq.\,(\ref{SER}), into the right hand side
of (\ref{K}) and using Eq.\,(\ref{TT}). Adding ${\cal K}$ to
the right hand side of Eq.\,(\ref{CAI}), we find
\begin{eqnarray}
\left . J_y^z\right |^I&=&\frac 1{4m\alpha} \sum_{{\bf p}\mu}\cos\phi_{\bf
p}(-1)^{\mu+1}\left \{{\rm Im}(\hat \Sigma_{1{\bf p}}^<)_{\mu\mu}
{\rm Im} (\hat {\rm G}_{0{\bf p}}^r)_{\mu\mu} -{\rm
Im}(\hat\Sigma_{0{\bf p}}^<)_{\mu\mu}
{\rm Im} (\hat {\rm G}_{1{\bf p}}^r)_{\mu\mu}\right \}\nonumber\\
&=&\sum_{\bf p}\frac {\cos\phi_{\bf p}}{4m\alpha} [{\rm Re} {\hat
I}_{11}-{\rm Re} {\hat I}_{22}].
\end{eqnarray}
Considering the diagonal parts of Eq.\,(\ref{EQTwo}), we finally
obtain
\begin{equation}
\sigma_{sH}^{I}\equiv \frac{\left . J_y^z\right |^I}{E}=-\sum_{\bf p}
\frac{ep_x}{4m\alpha p} \frac{\partial}{\partial p_x}
\left [f_1(p)-f_2(p)\right].\label{SSHH1}
\end{equation}

Although $\sigma_{sH}^{I}$ looks independent of the impurity
density, this spin-Hall conductivity arises essentially from a
disorder-mediated interband polarization. The longitudinal
transport of electrons driven by a dc electric field leads to
diagonal elements of the nonequilibrium distribution function
$\hat {\rm G}_{1{\bf p}}^<$, proportional to $n_i^{-1}$. Also,
these electrons participating in transport are scattered by
impurities to give rise to an interband polarization, which
becomes independent of the impurity density in the diffusive
regime. It is evident that the disorder plays an intermediate role
during such a polarization process.

When ignoring the collisional broadening in $\hat {\rm G}_{0{\bf
p}}^<$, distribution function $f_\mu (p)$ becomes the conventional
Fermi function, $f_\mu (p)\rightarrow n_{\rm
F}[\varepsilon_\mu(p)]$. The appearance of its derivative
$\partial f_\mu (p)/\partial p_x$, rather than $f_\mu (p)$ itself,
in Eq.\,(\ref{SSHH1}) implies that $\sigma_{sH}^I$ mainly relates
to the electron states near the Fermi surface. Note that this
point still remains reasonable even when considering a collisional
broadening in $\hat {\rm G}^r_{0{\bf p}}$ which is much smaller
than the Fermi energy.

\section{Results and discussions}

\subsection{Vanishing spin-Hall current in infinitely large Rashba 2D semiconductors}

We first consider the spin-Hall effect in an infinitely large
Rashba 2D semiconductor. In this case, the electron momentum is
continuous and the summation over the electron momentum in
Eq.\,(\ref{SSHH1}) can be replaced by a momentum integral.
Performing this momentum integral by parts, we find
$\sigma_{sH}^{I}=-\sigma_{sH}^{II}$, {\it i.e.} the total
spin-Hall conductivity vanishes.

Thus, we has analytically proven the vanishing of the total
spin-Hall current in infinitely large Rashba 2D electron systems
within the diffusive regime. Obviously, this elimination of the
spin-Hall current occurs quite generally: it is independent of the
specific form of scattering potential $V(p)$, of the impurity
density, of the SO coupling constant $\alpha$, and of temperature
$T$.

Ignoring the collisional broadening, $\sigma_{sH}^{II}$ becomes
independent of any scattering and takes the form
\begin{equation}
\sigma_{sH}^{II}=\frac{-e}{16\pi m\alpha}\int_0^{\infty}{\rm d} p
\{n_{\rm F}[\varepsilon_1(p)-\mu_c]-n_{\rm F}[\varepsilon_2(p)-\mu_c]\},
\end{equation}
with the chemical potential $\mu_c$. At zero temperature,
$\sigma_{sH}^{II}$ is equal to $e/8\pi$, in agreement with the
previous studies.\cite{t6,t9,t8} It is noted that in the case of
short-range scattering, there exists a simple relationship between
our result and the conclusion in the studies by means of Kubo
formula:\cite{t7,t9,t8} the $\sigma_{sH}^{II}$ and
$\sigma_{sH}^{I}$ in our treatment, respectively, correspond to
the bubble diagram and its vertex correction in the Kubo
formalism.

\subsection{Spin-Hall effect in finite-size Rashba 2D semiconductors}

\begin{figure}
\includegraphics [width=0.45\textwidth,clip] {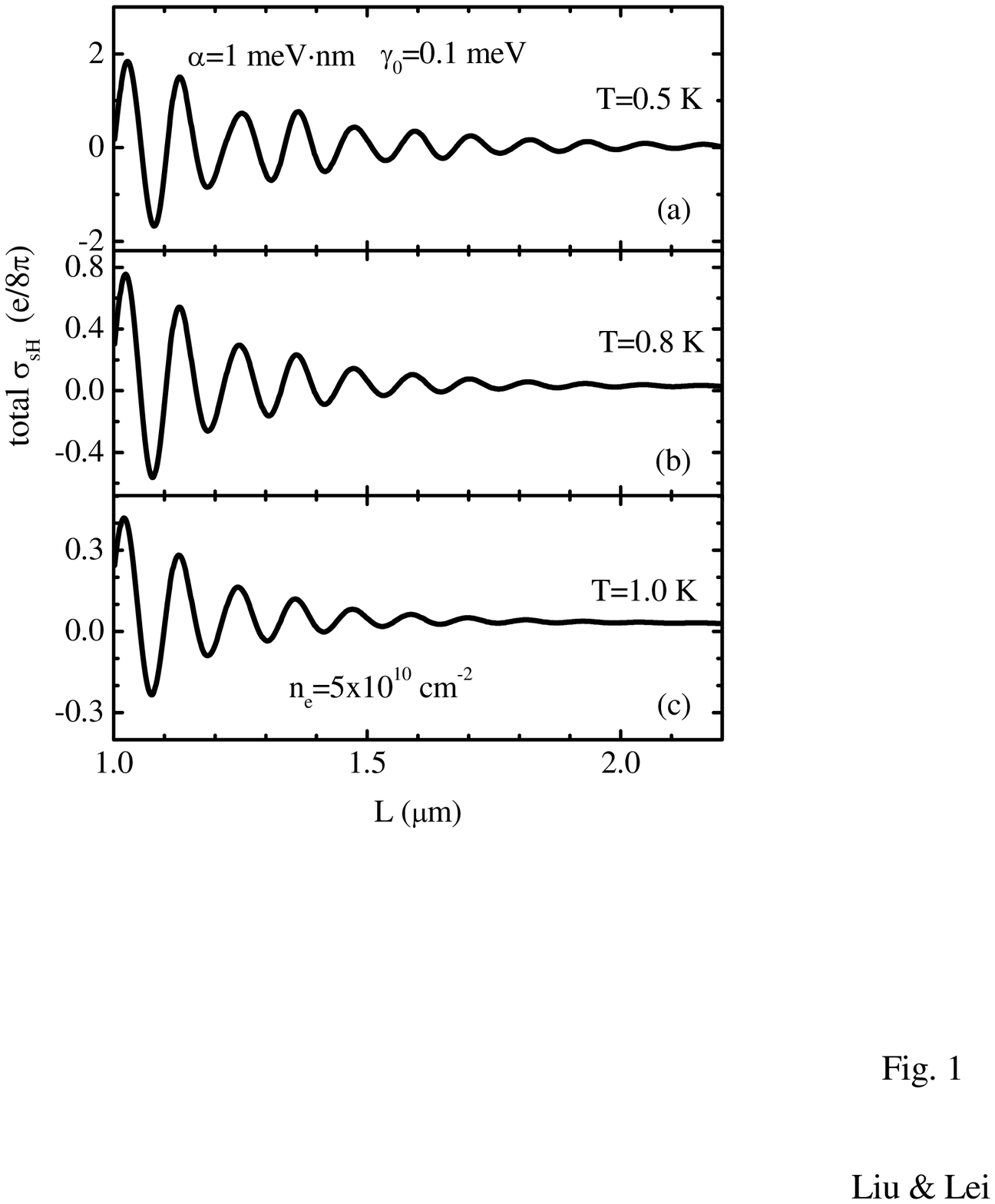}
\caption{Sample-size dependence of the total spin-Hall
conductivity $\sigma_{SH}$ in Rashba two-dimensional GaAs-based
semiconductors at different temperatures: (a) $T=0.5$\,K, (b)
$T=0.8$\,K, and (c) $T=1$\,K. The electron density and broadening
parameter are $n_e=5\times 10^{10}$\,cm$^{-2}$ and
$\gamma_0=0.1$\,meV. The SO coupling constant is
$\alpha=1$\,meV$\cdot$nm.} \label{fig1}
\end{figure}
\begin{figure}
\includegraphics [width=0.45\textwidth,clip] {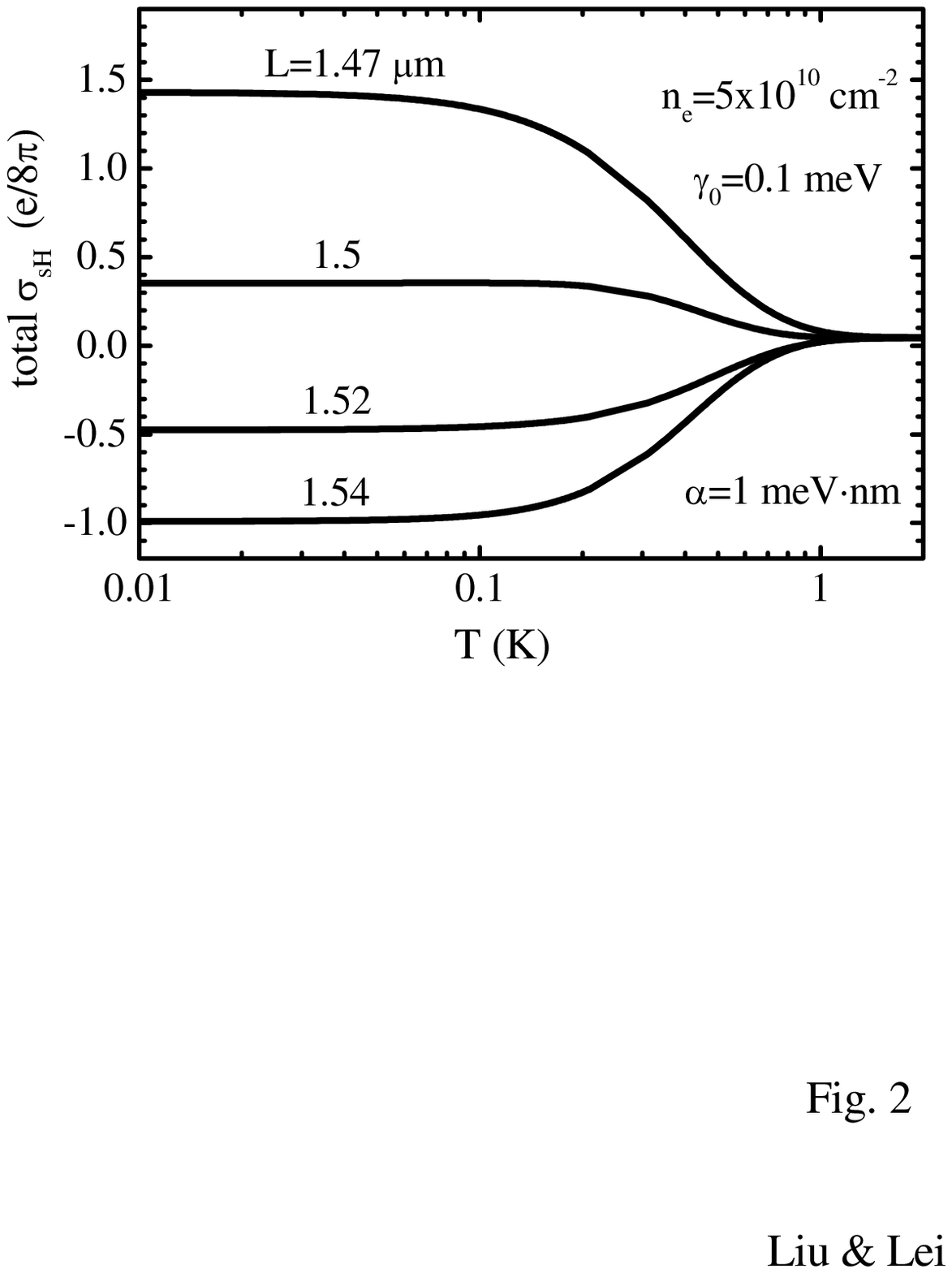}
\caption{Temperature dependence of the total spin-Hall conductivity
in 2D semiconductors of different sample sizes, $L=1.47$, $1.5$, $1.52$,
and $1.54$\,$\mu$m. Other parameters are the same as in Fig.\,1.}
\label{fig2}
\end{figure}
\begin{figure}
\includegraphics [width=0.45\textwidth,clip] {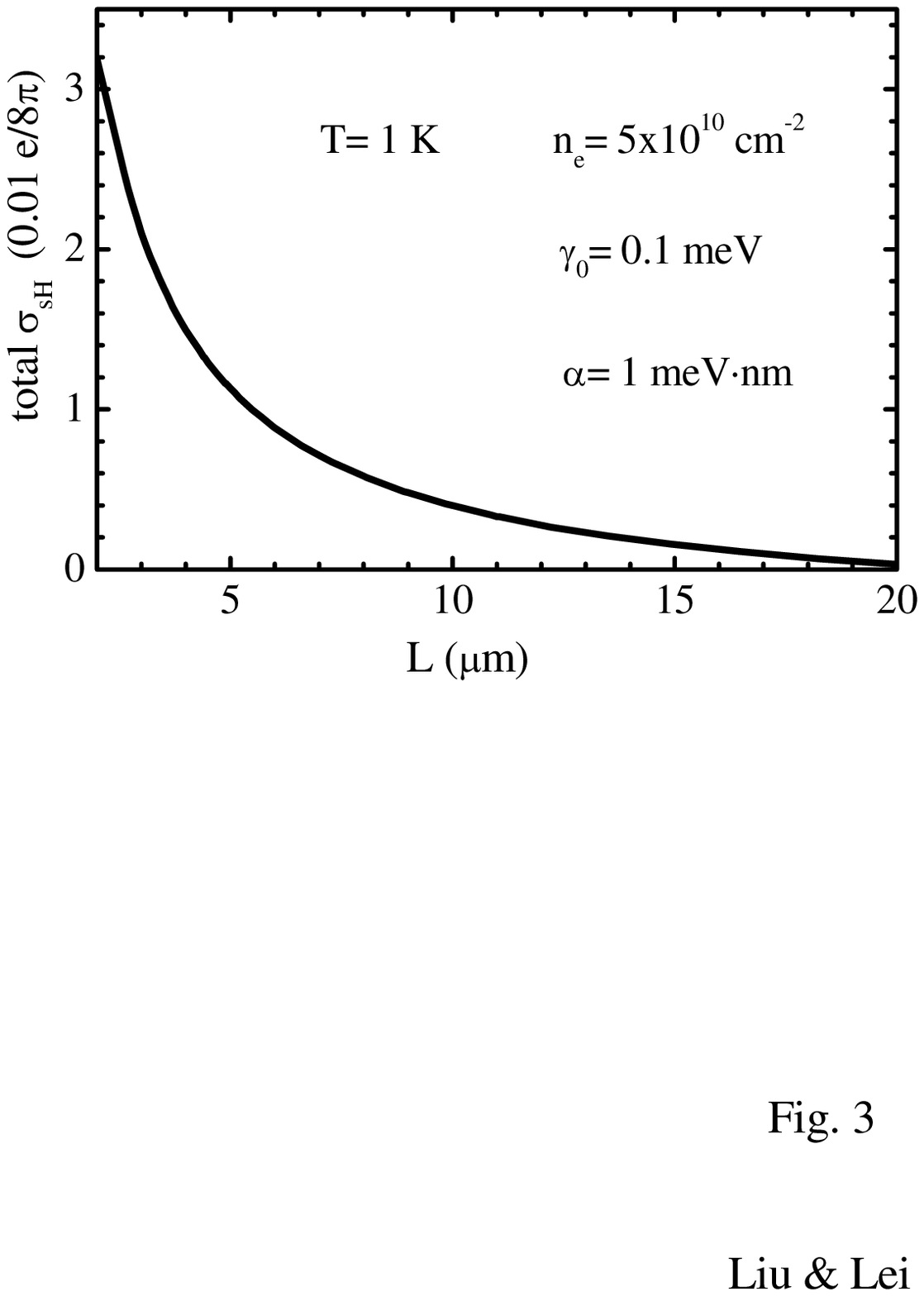}
\caption{Sample-size dependence of the total spin-Hall conductivity
in Rashba two-dimensional GaAs-based semiconductors at
temperature $T=1$\,K. Other parameters are the same as in Fig.\,1.}
\label{fig3}
\end{figure}

In the proof presented above, the summation over momentum was
replaced by a momentum integral. This is accurate only for samples
with large sizes. When the sample size is reduced to be comparable
with $2\pi/k_{F}$ [$k_{F}\equiv (k_{1{F}}+k_{2{F}})/2$ and
$k_{\mu{F}}$ is the Fermi momentum of electrons in spin-orbit
coupled (helicity) band $\mu$], the discretization of the electron
momentum can not be ignored. For definiteness, in present paper,
we consider a square Rashba 2D electron system with length $L$ in
both $x$ and $y$ directions. The possible values of the electron
momentum are  $p_x=2\pi n_x/L$ and $p_y=2\pi n_y/L$ with integers
$n_x$ and $n_y$. Since the disorder-mediated spin-Hall current is
associated mainly with the states near the Fermi surface while the
disorder-unrelated one is related to all electron states in the
Fermi sea, the discretization of the electron momentum has an
effect on $\sigma_{sH}^{I}$ stronger than that on
$\sigma_{sH}^{II}$. As a result, the total spin-Hall conductivity
may not be always vanishing in finite size samples.

In considering the effect of an energy (or momentum)
discretization, the collisional broadening of the retarded Green's
function should be taken into account. For this, we assume that
the imaginary part of the electron self-energy $\hat
{\Sigma}^r_{0{\bf p}}$ can be described by a constant parameter
$\gamma_0$: ${\rm Im}(\hat {\Sigma}^r_{0{\bf
p}})_{\mu\mu}=\gamma_0$. In this way, the distribution function
$f_\mu(p)$ takes a form, $f_\mu(p)={\rm
Im}[\Psi(1/2+C_\mu)]/\pi+1/2$, with the Digamma function  $\Psi
(x)$ and $C_\mu=[\gamma_0-i(\varepsilon_\mu-\mu_c)]/2\pi T$. [From
Eqs.\,(\ref{SSH1}) and (\ref{SSHH1}), we can see that the total
$\sigma_{sH}$ also vanishes in {\it infinitely large} samples
considering such a collisional broadening.] In present paper, we
restrict our discussion on the SHE in finite size samples within
the quasiclassical regime. In this regime, the sample size $L$ is
still much larger than $2\pi/k_F$ that a large number of electron
states are contained inside the Fermi surface (in the case
$L<2\pi/k_F$, the quantum size effect is important and the motion
of electrons will no longer be quasiclassical). Also, we restrict
our discussion to the diffusive regime: the impurities should be
enough dense that the electron mean free path $l=v_{F}\tau$
($v_{F}=k_F/m$ is the average Fermi velocity and $\tau$ is the
scattering time) is much less than $L$. Otherwise, the electron
motion will become ballistic. Under these considerations, all the
derivations in Sec. II, as well as Eqs.\,(\ref{SSH1}) and
(\ref{SSHH1}), remain valid for 2D systems of finite size.

Within the quasiclassical and diffusive regime, we have performed
a numerical study on the spin-Hall conductivity in a finite
GaAs-based heterojunction with a Rashba spin-orbit interaction. In
calculation, the SO coupling constant is chosen to be
$\alpha=1$\,meV$\cdot$nm. The electron density is $n_e=5\times
10^{10}$\,cm$^{-2}$, which indicates a Fermi wavevector
$k_F\approx 5.6\times 10^7$\,m$^{-1}$, a Fermi velocity
$v_F\approx 0.95\times 10^5$\,m/s, and a Fermi energy
$\epsilon_F\approx 1.8$\,meV (the electron effective mass of GaAs
is $m=0.068\,m_{e}$ with the free electron mass $m_e$). The
impurity density is assumed to give rise to a broadening parameter
$\gamma_0=0.1$\,meV, indicating a scattering $\tau\approx
3.3\times 10^{-12}$\,s, a mean free path $l\approx 0.3\,\mu$m, and
an electron mobility of order of $10$\,m$^2$/Vs. We consider only
the samples with sizes $L\geq 1$\,$\mu$m, which are much larger
than $2\pi/k_F$ and the mean free path $l$. The numerical results
obtained from Eqs.\,(\ref{SSH1}) and (\ref{SSHH1}) are plotted in
Figs.\,1, 2 and 3.

In Fig.\,1, the total spin-Hall conductivity
$\sigma_{sH}=\sigma_{sH}^I+\sigma_{sH}^{II}$, is shown as a
function of the sample size at three different temperatures
$T=0.5,0.8$ and 1.0\,K. We see that sensitively depending on the
size of micrometer samples, the spin-Hall conductivity can be
positive, zero and negative. At a given temperature, $\sigma_{sH}$
actually oscillates with a decreasing amplitude when increasing
the sample size from $L=1\,\mu$m. The period of the oscillation is
approximately equal to $2\pi/k_{F}$. Besides, there actually
exists another period, $2\pi/k_{{F}m}$ with $k_{{F}m}=(k_{1{
F}}-k_{2{F}})/2$. This period is quite large for the chosen
parameter $\alpha$ and its effect on $\sigma_{sH}$ becomes almost
unobservable. When temperature rises, the oscillation amplitude
decreases. Note that at low temperature, the maximum value of the
amplitude can be as large as $e/8\pi$.

In Fig.\,2, we plot the temperature dependence of the spin-Hall
conductivity at several different sample sizes, $L=1.47$, $1.5$,
$1.52$, and $1.54$\,$\mu$m (the spin-Hall conductivities for
samples of $L=1.47$\,$\mu$m and $L=1.54$\,$\mu$m, respectively,
correspond to a peak and a trough in Fig.\,1). At high
temperature, $\sigma_{sH}$ approaches a small (nonzero) constant
value. When temperature goes down from 1\,K, $\sigma_{sH}$ of
different $L$ spreads and approaches different values  between
$-e/8\pi$ and $1.4e/8\pi$. As a matter of fact, the finite-size
effect of $\sigma_{sH}$ originates from the rapid change of the
electron distribution around the Fermi surface, which is relevant
to three energy scales: (i) the finite-size induced energy
separation of the electron states around the Fermi surface,
$\Delta_F=2\pi v_F/L$, which is about $0.27$\,meV for
$L=1.5\,\mu$m, (ii) the collisional broadening of the energy level
described by the parameter $\gamma_0\approx 0.1$\,meV, and (iii)
the temperature $T$, which leads to a smearing of the distribution
function. Note that the finite size effect on $\sigma_{sH}$
remains nonvanishing when $\gamma_0$ and $T$ are smaller than
$\Delta_F$. When temperature $T$ increases from zero to
$\gamma_0$, the collisional broadening dominates the smeariness of
the electron distribution and the total spin-Hall conductivity
exhibits a plateau due to the temperature independence of
$\gamma_0$. As $T$ further increases to the range of $T>\gamma_0$,
the temperature smearing dominates and $|\sigma_{sH}|$ shrinks
with increasing $T$. When temperature becomes larger than
$\Delta_F$, $T>\Delta_F$, the effect of the energy level
separation is washed out and $\sigma_{sH}$ approaches a small
nonzero constant.

Note that the strong sample-size and temperature dependencies of
$\sigma_{sH}$ discussed above, come almost entirely from the
change of the disorder-mediated spin-Hall conductivity,
$\sigma_{sH}^I$, with variation of $L$ and $T$. Besides, there
exists another finite size effect arising mainly from the change
of the disorder-unrelated spin-Hall conductivity,
$\sigma_{sH}^{II}$, with sample size. It becomes important in the
larger $L$ scale, because in this case the finite size effect on
$\sigma_{sH}^I$ is washed out. In Fig.\,3, we plot the total
spin-Hall conductivity of finite Rashba two-dimensional GaAs-based
semiconductors having size from $L=2\,\mu$m to $L=20\,\mu$m at
temperature $T=1$\,K. We see that, though at this temperature the
$\sigma_{sH}$ oscillation disappears (Fig.\,1c) when $L>2\,\mu$m,
the spin-Hall conductivity remains to have a small finite value.
Only when the sample size increases to $L\geq 20\,\mu$m, can
$\sigma_{sH}$ close to zero, the result of an infinitely large
sample.

\section{Conclusions}

Employing a helicity-basis nonequilibrium Green's function
approach, we have investigated the spin-Hall effect in both the
infinitely large and finite-size Rashba two-dimensional electron
systems. A long-range electron-impurity scattering has been
considered in the self-consistent Born approximation. We found
that the spin-Hall effect originates from two different mechanisms
in helicity basis: disorder-unrelated and disorder-mediated
mechanisms. The disorder-unrelated mechanism corresponds to a
polarization process directly induced by dc electric field and is
associated with all electron states in the Fermi sea, while the
disorder-mediated one is the result of a polarization relating to
the nonequilibrium electrons participating in longitudinal
transport. In infinitely large diffusive Rashba 2D semiconductors,
the total spin-Hall current vanishes, independently of the
temperature, of the impurity density, of the specific form of the
isotropic scattering potential, and of the spin-orbit coupling
constant. However, when the sample size reduces, the spin-Hall
conductivity no longer always vanishes. Depending on the sample
size in the micrometer regime, the total $\sigma_{SH}$ can be
positive, zero or negative, with a maximum absolute value reaching
up to the order of magnitude of $e/8\pi$ at low temperatures. Such
a size effect shows up only at low temperatures. When temperature
increases that $T$ becomes comparable with the finite-size induced
energy separation of the electron states at the Fermi surface, the
$\sigma_{SH}$ oscillations disappear and spin-Hall conductivity
takes a small finite value before slowly approaching zero with
further increasing sample size to $L\geq 20\,\mu$m.

The present study indicates that a nonvanishing spin-Hall conductivity
may be obtained in a 2D Rashba electron systems of micrometer size,
notwithstanding its disappearance in infinitely large samples.
For it to appear, one has to accurately control the shape and size of
the sample. In addition, the mobility of the sample should be high
and the temperature should be low that both the collisional broadening of
the electron energy level and the temperature smearing are
smaller than the finite-size induced energy separation
of the electron states around the Fermi surface.

\begin{acknowledgments}
We would like to thank Drs. S.-Q. Shen and J. Sinova for useful
information. One of the authors (SYL) gratefully acknowledges
helpful discussions with Drs. W. S. Liu, Y. Chen and M. W. Wu.
This work was supported by projects of the National Science Foundation of
China, Shanghai Municipal Commission of Science
and Technology, and the Youth Scientific Research Startup Foundation of SJTU.
\end{acknowledgments}

{\it Note added}.---After our work was completed and submitted,
Adagideli and Bauer also reported the vanishing of spin-Hall
current in the presence of long-range scattering.\cite{Bauer}

\end{document}